\documentclass[aps,twocolumn,floatfix,altaffilletter,preprintnumbers,tightenlines,showpacs,showkeys,notitlepage,nofootinbib]{revtex4-2}

\usepackage[hidelinks,colorlinks,pdfusetitle]{hyperref}
\usepackage[normalem]{ulem}
\usepackage{amsmath,amssymb}
\usepackage{cancel}
\usepackage{relsize}
\usepackage{mathtools}
\usepackage{tabularx}
\usepackage{verbatim}
\usepackage{graphicx}               
\usepackage{upgreek}
\usepackage{url}
\usepackage{multirow}
\usepackage[dvipsnames]{xcolor}
\usepackage{float}

\definecolor{decay_purple}{rgb}{0.3,0,0.8}
\definecolor{ingoing}{HTML}{6395EE}
\definecolor{outgoing}{HTML}{FF5C5C}
\hypersetup{colorlinks=true,allcolors=decay_purple}

\usepackage{siunitx}
\usepackage{fontawesome}
\usepackage{enumitem}
\usepackage[capitalize]{cleveref}
\usepackage{lipsum}
\usepackage{gensymb}
\usepackage{booktabs}               
\usepackage{xspace}                 
\usepackage{pifont}
\usepackage{marvosym }
\usepackage{etoolbox,orcidlink}  

\usepackage{slashed}

\makeatletter

\renewcommand{\p@subsection}{}
\makeatother

\makeatletter
\def\l@subsubsection#1#2{}
\makeatother

\usepackage{tikz}
\usepackage{tikz-feynman}
\tikzfeynmanset{compat=1.1.0}
\makeatletter
\let\tikzfeynman@luatex@callback\relax
\makeatother

\usetikzlibrary{shapes,arrows,angles}
\usetikzlibrary{decorations.pathmorphing,decorations.markings}
\usetikzlibrary{positioning,arrows.meta}
\tikzfeynmanset{ fermion/.style = {
   decoration={
     markings,
     mark=at position 0.5
          with {\arrow[xshift=2mm]{Stealth[width=2mm,length=3mm]}}
     },
   postaction=decorate}
}

\tikzset{
  jline/.style={
    thick,
    postaction={decorate},
    decoration={markings, mark=at position 0.65 with {\arrow[scale=1.4]{Stealth}}}
  },
  vertex/.style={circle, fill=black, draw=black, inner sep=2.0pt},
  wavefront/.style={
    line width=0.3pt,
    decoration={snake, amplitude=0.5pt, segment length=3.5pt},
    decorate
  }
}

\newcommand{\Iint}[2]{%
  \mathcal{I}\!\left({\color{ingoing}#1},{\color{outgoing}#2}\right)%
}

\newcommand{\parenbar}[1]{\overset{
            \raisebox{-0.15em}{\scalebox{.4}{\textbf{(}}}
            \raisebox{-0.3em}{{\hspace{.03em}--\hspace{.05em}}}
            \raisebox{-0.15em}{\scalebox{.4}{\textbf{)}}}} {#1}}

\newcommand{\gsq}{\textcolor{gray}{\blacksquare}}
\newcommand{\bsq}{\blacksquare}

\newcommand{\bra}[1]{\ensuremath{\langle #1 |}}   
\newcommand{\ket}[1]{\ensuremath{| #1 \rangle}}   
\newcommand{\tr}{\text{tr}}

\begin{document}

\newcommand{\jgu}{PRISMA$^{++}$ Cluster of Excellence \& Mainz Institute for Theoretical Physics, \\
Johannes Gutenberg-Universit{\"a}t Mainz,
55099 Mainz, Germany}

\title{\textcolor{decay_purple}{Visible Neutrino Decay As An Open Quantum System}}

\author{Joachim Kopp \orcidlink{0000-0003-0600-4996}}
\email{jkopp@uni-mainz.de} 
\affiliation{\jgu}

\author{George A. Parker \orcidlink{0009-0000-1836-8696}}
\email{geparker@uni-mainz.de}
\affiliation{\jgu}

\begin{abstract}
{\centering\href{https://github.com/george-parker/nuDICE}{\large\faGithub}\\}
Decays of heavier neutrino mass eigenstates into lighter ones, while very slow in the Standard Model, can be significantly enhanced in scenarios with more than three neutrino flavours, or in models with new ultra-light particles such as Majorons. A full theoretical description is challenging due to the intricate interplay between oscillations and decay, interference between different decay channels, and the possibility of multi-step decay cascades. In this paper, we develop a fully general description of arbitrarily complex systems of oscillating and decaying neutrinos using methods from the theory of open quantum systems. Notably, we demonstrate how such systems can be implemented using the Lindblad master equation, the Liouvillian superoperator, as well as Kraus operators. The last two methods eschew the need for solving a differential equation, thereby showing superior numerical performance.
\end{abstract}

\maketitle

\section{Introduction}
\label{sec:intro}

Among all the unusual properties that neutrinos possess, their decay is perhaps the least studied one. This is not surprising, given that in the Standard Model, the rate for loop-suppressed radiative decays of the form $\nu_i \to \nu_j + \gamma$ is $\lesssim \SI{e-42}{yr^{-1}}$ \cite{Shrock:1974nd, Lee:1977tib, Marciano:1977wx, Pal:1981rm, Shrock:1982sc, Xing:2011zza}. (Here, $\nu_i$, $\nu_j$ denote neutrino mass eigenstates.) 

\begin{figure}[H]
    \centering
    \includegraphics[]{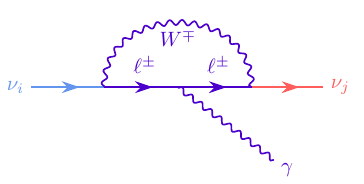}
\end{figure}

\noindent
However, neutrino decays can be significantly enhanced in extensions of the Standard Model that either contain new ultralight particles $\phi$, opening up new tree-level decay modes of the form $\nu_i \to \nu_j + \phi$ \cite{%
    Chikashige:1980ui, 
    Gelmini:1980re,    
    Valle:1983ua,      
    Porto-Silva:2020gma}, 
or that contain extra, heavier neutrino species.

\begin{figure}[H]
    \centering
    \includegraphics[]{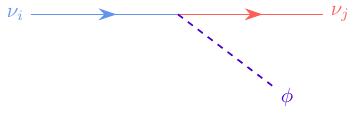}
\end{figure}

\noindent
The latter possibility is particularly interesting in the context of sterile neutrino dark matter, for which the non-observation of radiative decay provides some of the most important constraints \cite{%
    Dolgov:2000ew,     
    Abazajian:2001vt,  
    Drewes:2016upu,    
    Boyarsky:2018tvu,  
    Dasgupta:2021ies,  
    Krivonos:2024yvm}. 

Neutrino decay can be probed across a tapestry of scales and energies, with terrestrial experiments \cite{Abrahao:2015rba, Coloma:2017zpg, Gago:2017zzy, Choubey:2018cfz, Ternes:2024qui}, neutrino telescopes \cite{Maltoni:2008jr, Baerwald:2012kc, deSalas:2018kri} and cosmological observations \cite{Archidiacono:2013dua, Escudero:2019gfk, Barenboim:2020vrr}. The literature generally distinguishes between invisible decay,
\begin{align}
    \nu_i \to \textrm{?} ,
\end{align}
where all decay products are unobservable, depleting the overall neutrino number, and visible decay
\begin{align}
    \nu_i \to \nu_j + \phi ,
\end{align}
where the final state includes a detectable neutrino \cite{Lindner:2001fx, Gago:2017zzy, Porto-Silva:2020gma, Abdullahi:2020rge}.

The theoretical description of neutrino decay is complicated by the intricate interplay between oscillations and decay, interference between different decay channels, and the possibility of multi-step decay cascades of the form $\nu_i \to \nu_j \to \nu_k$. Therefore, many past works have made simplifying assumptions (absence of some decay modes, neglecting certain interference effects). Without such approximations, the formalism quickly becomes unwieldy \cite{Lindner:2001fx}, especially when there are more than three neutrino species.

In this paper, we therefore take a different approach: leveraging techniques from quantum information, we develop open quantum systems descriptions of neutrino decay. In particular, we will treat the problem using the Lindblad master equation, the Liouvillian superoperator, and Kraus operators. The latter methods will allow us to write the state of the combined oscillation+decay system at any time during its evolution without the need for solving a differential equation. The method works for arbitrary numbers of neutrino species and decay modes, and it keeps track of the full energy spectrum of the neutrino ensemble.

Some steps in a similar direction have been taken in the past in ref.~\cite{Stankevich:2024xyc} (see also ref.~\cite{Stankevich:2020icp} for related work). We also emphasize the similarity between a system of oscillating and decaying neutrinos with a system of neutral mesons ($K^0$, $D^0$, $B^0$). The evolution of these systems has been described with an open quantum systems formalism in refs.~\cite{Bernabeu:2012au, Caban:2005ue, Bertlmann:2006fn}. Moreover, the open quantum systems formalism is commonly used for studying neutrino quantum decoherence absorption \cite{%
    Benatti:2000ph,    
    Lisi:2000zt,       
    Gago:2000qc,       
    Ohlsson:2000mj,    
    Oliveira:2016asf,  
    Coelho:2017byq,    
    Coloma:2018idr,    
    Nieves:2020jjg,    
    DeRomeri:2023dht}, 
which could arise from wave-packet separation \cite{Akhmedov:2022bjs}, production and detection uncertainties \cite{Ohlsson:2000mj}, or even propagation through quantum gravity fluctuations \cite{Alexandre:2007na, Stuttard:2020qfv}. 

This work is structured as follows: in \cref{sec:existing-results}, we summarize the relevant expressions for the neutrino decay rates and we review the phenomenological formalism from ref.~\cite{Lindner:2001fx}. In \cref{sec:oqs} we then introduce the formalism of open quantum systems, before applying it to neutrinos in \cref{sec:neutrino-decay-as-oqs}. We compare the different formalisms in \cref{sec:discussion}, before concluding in \cref{sec:conclusions}. A Python implementation of the methods developed in this paper is available on GitHub \cite{github}.

\section{Existing Results on Neutrino Decay}
\label{sec:existing-results}

\subsection{Decay Rates}
\label{sec:decay}

While neutrino decay occurs even in the Standard Model ($\nu_j \to \nu_k + \gamma$), the corresponding rates are too small to be phenomenologically relevant. This changes in the presence of heavier, sterile, neutrinos or in the presence of additional light particles that open up new decay modes. In the following, we will assume the latter situation, and we will, for definiteness, work in a simple model featuring a massless (or very light) Majoron ${J}$ with the following Lagrangian
\begin{align}
    \mathcal{L} \supset \frac{g^s_{ij}}{2} \bar{\nu}_j \nu_i {J}
        + \frac{g^p_{ij}}{2} \bar{\nu}_j i \gamma_5 \nu_i J
        + \text{h.c.} ,
\end{align}
where $g^s_{ij}$ and $g^p_{ij}$ are the Majoron's dimensionless scalar and pseudoscalar coupling constants, respectively. If neutrinos are Dirac particles, only lepton number conserving processes ($\nu_i \to \nu_j + J^\dag$, $\bar\nu_i \to \bar\nu_j + J$) can occur \cite{Abdullahi:2020rge, Kim:1990km}, while for Majorana neutrinos, also lepton number violating processes like $\nu_i \to \bar\nu_j + J$ are possible.

The partial widths for the decay of a parent neutrino of energy $E_i$ into a daughter neutrino of energy $E_j$ in the lab frame are \cite{Kim:1990km}
\begin{align}
    \Gamma_{ij}^\text{C}
        &= \frac{m_i m_j}{16 \pi E_i}
           \left[(g^s_{ij})^2\,\mathfrak{f}_{ij}
             +   (g^p_{ij})^2\,\mathfrak{h}_{ij} \right], \\
    \Gamma_{ij}^\text{V}
        &= \frac{m_i m_j}{16 \pi E_i}
           \left[\left((g^s_{ij})^2 + (g^p_{ij})^2 \right)
                 \mathfrak{K}_{ij} \right],
\end{align}
where the superscript C (V) stands for lepton number conserving (violating) modes, and we define the kinematic functions
\begin{align}
    \mathfrak{f}(x_{ij}) &= \frac{x_{ij}}{2}
                          + 2
                          + \frac{2}{x_{ij}} \log x_{ij}
                          - \frac{2}{x_{ij}^2}
                          - \frac{1}{2 x_{ij}^3}, \\
    \mathfrak{h}(x_{ij}) &= \frac{x_{ij}}{2}
                          - 2
                          + \frac{2}{x_{ij}} \log x_{ij}
                          + \frac{2}{x_{ij}^2}
                          - \frac{1}{2 x_{ij}^3}, \\
    \mathfrak{K}(x_{ij}) &= \frac{x_{ij}}{2}
                          - \frac{2}{x_{ij}} \log x_{ij}
                          - \frac{1}{2 x_{ij}^3}.
\end{align}
with the shorthand notation $x_{ij} \equiv m_i/m_j$. The corresponding differential decay rates are\footnote{Note that our definition of $\eta_{ij}$ differs from the one in ref.~\cite{Lindner:2001fx}, where $\eta_{ij}=\frac{1}{\Gamma_{ij}}\frac{d\Gamma_{ij}}{dE_j}$.}
\begin{align}
    \begin{split}
        \eta_{ij}^\text{C}(E_i,E_j) &\equiv \frac{d\Gamma_{ij}^\text{C}}{dE_j}
                                            \notag\\
            &= \frac{m_i m_j}{16 \pi E_i^2}
               \left[(g^s_{ij})^{2} \left(\mathfrak{a}_{ij} + 2 \right)
                    +(g^p_{ij})^{2} \left(\mathfrak{a}_{ij} - 2 \right) \right]
                \hat\Theta, \\
        \eta_{ij}^\text{V}(E_i,E_j) &\equiv \frac{d\Gamma_{ij}^\text{V}}{dE_j}
                                            \notag\\
            &= \frac{m_i m_j}{16 \pi E_i^2}
               \left[(g^s_{ij})^2 + (g^p_{ij})^{2}\right]
               \left( \frac{1}{x_{ij}} + x_{ij} - \mathfrak{a}_{ij} \right)
               \hat\Theta,
    \end{split}
    \label{eq:dwid}
\end{align}
with
\begin{align}
    \mathfrak{a}_{ij} &= \frac{1}{x_{ij}} \frac{E_i}{E_j}
                       + x_{ij} \frac{E_j}{E_i}.
\end{align}
We enforce the kinematic constraint
\begin{align}
    \frac{E_i}{x_{ij}^2} \leq E_j \leq E_i \, ,
\end{align}
via the Heaviside functions
\begin{align}
    \hat{\Theta} \equiv \Theta\big( E_j - E_i/x^2_{ij} \big) \, \Theta\big( E_i - E_j \big).
    \label{eq:heaviside}
\end{align}
In the following, we will for simplicity set $g^s_{ij} = g^p_{ij} \equiv g_{ij}$. Moreover, unless otherwise indicated, we will assume neutrinos to be Dirac particles.

\subsection{Phenomenological Approach to Neutrino Decay and Oscillation}
\label{sec:owl-approach}

\begin{figure*}
    \centering
    \includegraphics[]{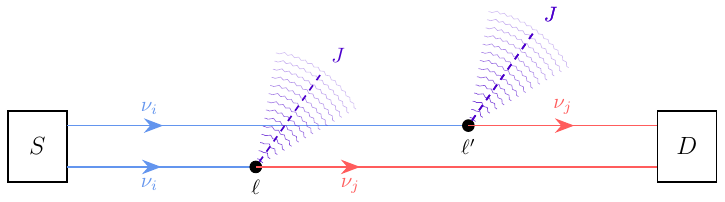}
    \caption{Neutrino production from the source ($S$), propagation, and absorption in a detector ($D$). We show two neutrinos decaying at different positions along the trajectory, and we illustrate the outgoing Majoron wave in each case. Due to the different Majoron kinematics, the two paths are distinguishable and cannot interfere, as discussed below \cref{eq:pheno}.}
    \label{fig:propagation-diagram}
\end{figure*}

Most discussions of neutrino decay in the literature take a phenomenological perspective, starting from the Hamiltonian describing standard three-flavour neutrino oscillations. In the mass basis, it is given by
\begin{equation}
    H = \frac{1}{2E} \begin{pmatrix}
                         m^2_1 &               & \\
                         & m^2_2  & \\
                         &                 & m^2_3
                     \end{pmatrix} \cong
        \frac{1}{2E} \begin{pmatrix}
                         0 &               & \\
                           & \Delta m_{21}^2 & \\
                           &                 & \Delta m_{31}^2
                     \end{pmatrix},
    \label{eq:ham}
\end{equation}
with $E$ the neutrino energy and $\Delta m_{jk}^2 \equiv m_j^2 - m_k^2$. The $\cong$ symbol indicates that the matrix on the right is equivalent to the one on the left up to terms proportional to the identity matrix; such terms contribute only a global phase to the evolution and can therefore be dropped. In the ultra-relativistic limit, the $\nu_\alpha \to \nu_\beta$ flavour oscillation probability without decay for neutrinos of energy $E$ is
\begin{equation}
    P_{\alpha\beta}(E,L)
        = \bigg| \sum_k U_{\alpha k}^{*} U_{\beta k}
          \exp\Big( -i\frac{m_k^2 L}{2E} \Big) \bigg|^2,
    \label{eq:Pab}
\end{equation}
where $U_{\alpha k}$ are the elements of the leptonic mixing matrix and $L$ is the distance the neutrinos travel.

It is straightforward to introduce \emph{invisible} neutrino decay by adding an imaginary part to the Hamiltonian, rendering it non-Hermitian,
\begin{align}
    H^\mathrm{invis} \equiv H - i {\Gamma}.
    \label{eq:no-herm}
\end{align}
The decay matrix, $\Gamma$, is a real $3 \times 3$ matrix which is often chosen to be diagonal in the neutrino mass basis. We make the same assumption here, but we emphasize that off-diagonal decay matrices are possible and that the results of this paper can be applied in this more general case as well \cite{Berryman:2014yoa, Chattopadhyay:2021eba, Chattopadhyay:2022ftv}. Assuming that $m_1$ is stable, $\Gamma$ takes the form
\begin{equation}
    {\Gamma} = \frac{1}{2E} \begin{pmatrix}
                                0 &          & \\
                                  & \alpha_2 & \\
                                  &          & \alpha_3
                            \end{pmatrix}.
    \label{eq:gam_matrix}
\end{equation}
where we have introduced the energy-independent and Lorentz-invariant decay parameter $\alpha_i = E \sum_j \Gamma_{ij}(E)$.

When $[H, \Gamma] = 0$, so that the two matrices are diagonal simultaneously, the effect of $\Gamma$ is to add an exponential damping term to \cref{eq:Pab}. Defining the oscillation+decay phase, $\upphi_i$, as
\begin{align}
  \upphi_i(E) = \frac{im_i^2}{2 E}+\frac{\alpha_i}{2 E} ,
\end{align}
the probability for oscillation plus invisible decay is
\begin{align}
    P^\mathrm{invis}_{\alpha\beta}(E, L) =
        \bigg| \sum_k U_{\alpha k}^* U_{\beta k}
        \exp\big[-\upphi_k(E) L \big] \bigg|^2 .
    \label{eq:Pab_inv}
\end{align}

Turning to visible decays, $\nu_i \to \nu_j + \phi$, the formalism becomes significantly more complex as both the disappearance of parent neutrinos and the appearance of daughter neutrinos need to be described. Moreover, we now need to carefully distinguish between the parent energy $E$ and the daughter energy $E'$. In practice, the appearance of daughter neutrinos is usually described by adding a ``regeneration'' term $\mathcal{R}_{\alpha\beta}(E',L)$ to the probability of invisible decay, \cref{eq:Pab_inv}:
\begin{align}
    P^\mathrm{vis}_{\alpha\beta}(E',L)
        = P^\mathrm{invis}_{\alpha\beta}(E',L) + \mathcal{R}_{\alpha\beta}(E',L) .
        \label{eq:Pab_vis}
\end{align}
This approach has been discussed in detail by Ohlsson, Winter, and Lindner in ref.~\cite{Lindner:2001fx}, so we denote it in the following as OWL. \Cref{eq:Pab_vis} describes the probability of a neutrino with initial flavour $\alpha$ and initial energy $E$ to oscillate or decay to a daughter neutrino of flavour $\beta$ at any energy. The regeneration term is given by
\begin{align}
     \mathcal{R}_{\alpha\beta}(E',L) =
         \mathlarger{\int}_0^\infty \! dE \,
         \mathlarger{\int}_0^L \! dl \, \Big| \sum_{i>j} U_{\alpha i}^* U_{\beta j}\, 
        \mathcal{A}_{ij}(E, E') \Big|^2 ,
    \label{eq:regen}
\end{align}
where the first integral runs over the energy of the daughter neutrino, $L$ is the baseline, and $\mathcal{A}_{ij}$ is the $\nu_i \to \nu_j$ oscillation+decay amplitude given by \cite{Lindner:2001fx, Abdullahi:2020rge}
\begin{align}
    \mathcal{A}_{ij}(E,E') =
            \sqrt{\eta_{ij}(E, E')}
            \,\exp\big[ -\upphi_i(E) \, l
              - \upphi_j(E') (L-l)  \big] .
    \label{eq:pheno}
\end{align}
Here the first term in the exponent, $-\upphi_i(E) \, l$, describes the evolution of the parent neutrino before it decays at distance $l$, whilst the second term, $-\upphi_j(E') (L-l)$, describes the phase accumulation of the daughter state up to detection at distance $L$. Note that the energy integral in \cref{eq:pheno} run from 0 to $\infty$ as the kinematic constraints are handled by the Heaviside factors in \cref{eq:dwid}.

The integral over the decay vertex, $l$, in \cref{eq:Pab_vis} is carried out at the probability level, not at the amplitude level. This is because decays occurring at different $l$ are distinguishable due to the different Majoron kinematics \cite{Lindner:2001fx}. We illustrate this in \cref{fig:propagation-diagram}.

We have assumed here at most one decay over distance $L$; for cascade decays, e.g.\ $\nu_3 \to \nu_2 \to \nu_1$,  we would need to evaluate multiple nested integrals of the form $\int_0^L dl_1 \int_{l_1}^L dl_2 \cdots \int_{dl_{n-1}}^L dl_n$, with $n$ amplitudes $\mathcal{A}_{ij}$ appearing in the integrand.

\section{Open Quantum Systems}
\label{sec:oqs}

We now introduce the concepts from the theory of open quantum systems which we will use to reformulate the neutrino oscillation+decay problem.

\subsection{Toy Example}
\label{sec:toy-example}

\begin{figure}[b]
    \centering
    \includegraphics[]{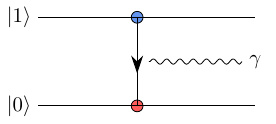}
    \caption{Schematic diagram of the excited state of a two-level atom decaying via spontaneous photon emission.}
    \label{fig:ampdamp}
\end{figure}

To set the stage and introduce the basic idea, we consider as a toy example the \textit{amplitude damping} model of an excited state of an atom ($|1\rangle$) decaying to the ground state ($|0\rangle$) via spontaneous emission of a photon. The quantum state of such a system, or of an ensemble of such systems, can be described by a $2 \times 2$ density matrix, $\rho$, which obeys the equation
\begin{align}
    \frac{d\rho}{dt} = L^\dag \rho L - \frac{1}{2}\{ L^\dag L, \rho \},
    \quad\text{with}\quad
    L = \begin{pmatrix}
            0 & \sqrt{\Gamma} \\
            0 & 0
        \end{pmatrix}.
    \label{eq:lindblad-toy}
\end{align}
This is a special case of the \emph{Lindblad master equation}, which we will disscuss further below. Alternatively, time evolution of $\rho$ can be written as
\begin{align}
   \rho(t) = M_0 \rho M_0^\dag + M_1 \rho M_1^\dag ,
   \label{eq:kraus-evolution-toy}
\end{align}
where the \emph{Kraus operators} are
\begin{align}
    \begin{split}
        M_0(t) &= \begin{pmatrix}
                      1 & 0 \\
                      0 & \sqrt{e^{-\Gamma t}}
                  \end{pmatrix}, \\
        M_1(t) &= \begin{pmatrix}
                      0 & \sqrt{1-e^{-\Gamma t}} \\
                      0 & 0
                  \end{pmatrix}.
    \end{split}
    \label{eq:kraus-operators-toy}
\end{align}
In the Lindblad formulation, \cref{eq:lindblad-toy}, it is usually relatively straightforward to implement the desired physical phenomena (unitary evolution, decay, decoherence, etc.), but the approach still requires solving a differential equation. The use of Kraus operators directly gives the solution of the time evolution, but determining the Kraus operators is more involved for complex systems.

\subsection{General Formalism}
\label{sec:general-formalism}

We now discuss in more general terms the formalism introduced in the toy example. Our starting point is again a density matrix, $\rho$, describing a quantum system, for instance an ensemble of neutrinos. An operation which describes unitary time evolution can be represented by a unitary operator, $U$, acting on $\rho$:
\begin{align}
    \rho \to U \rho U^{\dagger}.
    \label{eq:UrhoU}
\end{align}
The operator $U$ can be determined based on a Hamiltonian $H$ by solving the von Neumann equation
\begin{align}
    \frac{d\rho}{dt} = -i \left[H, \rho \right], 
    \label{eq:von-Neumann}
\end{align}
with the familiar solution $\rho(t) = e^{-i H t}\rho(0)e^{i H t}$.

Going beyond unitary dynamics, \cref{eq:von-Neumann} can be generalized to the Gorini--Kossakowski--Sudarshan--Lindblad master equation (or Lindblad master equation, or simply \emph{Lindbladian}), which allows for the description of non-unitary dynamics such as decoherence, damping, and decay. The equation reads
\begin{align}
    \frac{d\rho}{dt}= -i\left[H, \rho\right] - \mathcal{D}(\rho) ,
    \label{eq:lindblad-general}
\end{align}
with the \emph{dissipator}, $\mathcal{D}(\rho)$. The dissipator has to respect the principles of trace preservation and complete positivity, which restricts it to the form \cite{Gorini:1975nb, Lindblad:1975ef}
\begin{align}
    \mathcal{D}(\rho) = \sum_{k} \Big[ L_k \rho L_k^\dagger
        - \frac{1}{2}\{L_k^\dagger{L_k}, \rho\} \Big].
    \label{eq:dissipator}
\end{align}
The $L_k$ are called \emph{Lindblad operators}; each of them encodes a particular non-unitary operation, depending on the details of the physical system. In \cref{eq:lindblad-toy}, we have seen a concrete example. The Lindblad master equation (or ``Lindbladian''), \cref{eq:lindblad-general} is the first of three representations of an open quantum system that we discuss in this section. For more information on these representations and how to convert between them, see \cite{Havel:2003zne,Wood:2011zvw}.

The second representation is based on a vectorized version of \cref{eq:lindblad-general}, called the \emph{unravelled} master equation. Using the vectorization operation, $\operatorname{vec}$, which stacks the columns of a matrix, from left to right, into a vector, the unravelled master equation reads
\begin{align}
    \frac{d}{dt}\operatorname{vec}(\rho)=\hat{\mathcal{L}} \operatorname{vec}(\rho).
    \label{eq:vec}
\end{align}
where $\hat{\mathcal{L}}$ is called the \emph{Liouvillian superoperator}, or simply the Liouvillian. $\hat{\mathcal{L}}$ is an $n^2 \times n^2$ matrix (or, equivalently, a rank-4 tensor) given by
\begin{multline}
    \hat{\mathcal{L}} =
        -i \left( \mathbb{I} \otimes H
             - H^{\mathrm{T}} \otimes \mathbb{I}\right)
  + \sum_k L_k^* \otimes L_k \\
  - \frac{1}{2} \mathbb{I} \otimes L_k^{\dagger} L_k
  - \frac{1}{2}(L_k^{\dagger} L_k)^{\mathrm{T}} \otimes \mathbb{I} ,
    \label{eq:ume}
\end{multline}
where $\otimes$ denotes the Kronecker product. This leads to the formal solution
\begin{align}
    \operatorname{vec}[\rho(t)] = \mathcal{E}(t)\, \operatorname{vec}[\rho(0)], \qquad \mathcal{E}(L) = e^{\hat{\mathcal{L}} t},
    \label{eq:t-evolution-dynamical-map}
\end{align}
where $\mathcal{E}(t)$ is called the \emph{dynamical map}. Like $\hat{\mathcal{L}}$, also $\mathcal{E}(t)$ is an $n^2 \times n^2$ matrix, or equivalently a rank-4 tensor $\mathcal{E}_{mn,\mu\nu}$. Reshuffling its entries yields the \emph{Choi matrix} \cite{Choi:1975nug}, $\mathcal{C}$, in the following way \cite{Jamiolkowski:1972pzh}:
\begin{align}
    \mathcal{C}_{\nu n, \mu m} \equiv \mathcal{E}_{mn,\mu\nu}
    \label{eq:shuffle}
\end{align}
Consider now the spectral decomposition of $\mathcal{C}$,
\begin{align}
    \mathcal{C} = \sum_k \lambda_k \, \ket{\phi_k} \bra{\phi_k},
    \label{eq:Choi-decomposition}
\end{align}
where $\lambda_k$ and $\ket{\phi_k}$ are the eigenvalues and eigenvectors of $\mathcal{C}$, respectively. In component notation, \cref{eq:Choi-decomposition} reads
\begin{align}
    \mathcal{C}_{\nu n,\mu m} = \sum_k \lambda_k \, \phi_{k,\nu n} \phi^*_{k,\mu m} = \mathcal{E}_{mn,\mu\nu} .
    \label{eq:Choi-decomposition-components}
\end{align}
Inserting this into \cref{eq:t-evolution-dynamical-map}, the time evolution of $\rho$ can be written as
\begin{align}
    \rho_{mn}(t) &= \mathcal{E}_{mn,\mu\nu}(t) \, \rho_{\mu\nu}(0) \notag\\
                 &= \sum_k \lambda_k \, \phi^{(k)}_{\nu n}(t) \, \phi^{(k)*}_{\mu m}(t) \,\rho_{\mu\nu}(0) \,.
    \label{eq:t-evolution-kraus1}
\end{align}
This suggests the definition of \emph{Kraus operators}, $M_k$, according to
\begin{align}
    M_{k, m \mu} \equiv \sqrt{\lambda_k} \phi^*_{k,\mu m} \,.
    \label{eq:kraus-op-definition}
\end{align}
The time-evolution of the system can be written in terms of Kraus operators as 
\begin{align}
    \rho(t) = \sum_k M_k(t) \rho(0) M_k^\dag(t) .
    \label{eq:t-evolution-kraus2}
\end{align}
This is the third and final representation of an open quantum system we will use in this paper. Note that the Kraus operators have to satisfy the completeness relation\footnote{This follows from the fact that the time-evolution of the system has to preserve the trace of the density matrix, that is,
\begin{align}
    \tr\Big( \sum_k M_k(t) \rho(0) M_k^\dag(t) \Big) = \tr\rho(0).
\end{align}
Using the cyclicity of the trace, and the fact that the relation has to hold for arbitrary $\rho(0)$, the completeness relation follows.
}
\begin{align}
    \sum_k M_k^\dag M_k = \mathbb{I} .
    \label{eq:Kraus-completeness}
\end{align}

\section{Neutrino Decay as an Open Quantum Systems}
\label{sec:neutrino-decay-as-oqs}

\subsection{Lindblad Approach}
\label{sec:lindblad-approach}

In this section, we pose the visible neutrino decay system as an open quantum system, deriving the main results of the present work.

Some steps in this direction have been taken in the existing literature by invoking a modified von Neumann equation of the form \cite{Moss:2017pur, Dentler:2019dhz}
\begin{align}
    \frac{d\rho(E)}{dt} = -i \left[H(E), \rho\right]
        - \left\{ \Gamma(E), \rho \right\} + \mathcal{R}(E),
    \label{eq:vN_decay}
\end{align}
where the first term describes standard oscillations, the second, anti-Hermitian, term describes disappearance of parent neutrinos, and the third term is a regeneration term describing the creation of daughter neutrinos \cite{Moss:2017pur, Dentler:2019dhz}. Note the similarity between \cref{eq:vN_decay,eq:lindblad-general} (the factor of $\frac{1}{2}$ from \cref{eq:dissipator} has been absorbed into the definition of $\Gamma(E)$).

This approach has been implemented in the \texttt{nuSQuIDSDecay} package \cite{Delgado:2014lyt, Arguelles:2021twb, Moss:2017pur}, which provides an integro-differential equation solver to time-evolve an ensemble of neutrinos distributed across $N$ discrete energy bins. The neutrinos in each bin are described by a $3\times3$ density matrix, $\rho(E)$. However, \texttt{nuSQuIDSDecay} does not account for the interference between multiple neutrino decay modes.  Including such interference effects is one of the ways in which the present study goes beyond the existing literature.

Our starting point is the Lindblad master equation, cf.\ \cref{eq:lindblad-general,eq:dissipator}. Neutrino decay has been studied in this way before in ref.~\cite{Stankevich:2024xyc} (see also ref.~\cite{Stankevich:2020icp} for related work), albeit without considering the energy spectrum of daughter neutrinos and without solving it with more than one nonzero decay mode. The Lindblad formalism is also commonly used to describe neutrino decoherence and absorption \cite{%
    Benatti:2000ph,    
    Lisi:2000zt,       
    Gago:2000qc,       
    Ohlsson:2000mj,    
    Oliveira:2016asf,  
    Coelho:2017byq,    
    Coloma:2018idr,    
    Nieves:2020jjg,    
    DeRomeri:2023dht}. 

As we are interested in the energy spectrum of the final state neutrino ensemble, we split the neutrino ensemble into $N_E$ energy bins, so that $\rho$ should be understood as a $(3 N_E) \times (3 N_E)$ matrix, or more generally as an $(N_\nu N_E) \times (N_\nu N_E)$ matrix if we consider $N_\nu > 3$ neutrino species. Note, however, that only the diagonal $N_\nu \times N_\nu$ blocks are non-zero. Elements outside these blocks correspond to coherences between neutrinos of different energies. But any such correlation decoheres almost instantaneously as it comes with a fast oscillating phase factor of the form $e^{\delta\!E\, t}$, where $\delta E$ is the (macroscopic) energy difference between the two states. When integrating over a macroscopic energy interval (the width of our energy bins, the detector resolution), these phase factors average to zero. Therefore, it is in practice only necessary solve a system of dimension $N_E \times N_\nu \times N_\nu$, not a system of size $(N_\nu N_E) \times (N_\nu N_E)$.

We now discuss the form of the Lindblad operators we need to introduce in \cref{eq:dissipator} to describe neutrino decay. Na\"ively, we might consider introducing one operator per decay mode $(i,n) \to (j,m)$, where the indices $i$, $j$ label the initial and final neutrino mass eigenstates ($i > j$), and $n$, $m$ label the initial and final energy bins. However, this set of operators would neglect correlations between daughter neutrinos of the same energy, but different mass. For instance, when $\nu_3$ can decay to both $\nu_2$ and $\nu_1$, each $\nu_3$ effectively decays to a well-defined \emph{superposition} of $\nu_1$ and $\nu_2$.\footnote{Note that such correlations exist only between daughter states of the same energy. Daughters with different energy are distinguishable as discussed above.} Therefore, decays originating from the same parent state $(i,n)$ and going to the same final state energy bin $m$ should be combined into a single Lindblad operators. The Lindblad equation now takes the form
\begin{multline}
    \frac{d\rho^{(m)}}{dt}= - i\big[H^{(m)}, \rho^{(m)} \big]
                            - \sum_{i,n} \Big(
                                  L_i^{(nm)} \rho^{(n)} L_i^{(nm)\dag} \\
                                - \frac{1}{2}\big\{L_i^{(nm)\dag} L_i^{(nm)}, \; \rho^{(m)}\big\}
                              \Big) ,
    \label{eq:lindblad-neutrino}
\end{multline}
where $\rho^{(m)}$ denotes the $m$-th diagonal block of $\rho$ and the Lindblad operators on the right-hand side are
\begin{align}
    L_i^{(nm)} = \sum_j \upgamma_{ij}^{(nm)} \; \ket{j,m} \bra{i,n}
    \label{eq:lindop}
\intertext{with}
    [\upgamma_{ij}^{(nm)}]^2 = \int_{E_m^-}^{E_m^+}
    \eta_{ij}(E_n, E')\, dE' .
    \label{eq:gamma-ijnm}
\end{align}
Here, $E_m^\pm$ are the upper and lower edges of the $m$-th energy bin; kinematic constraints are accounted for by the Heaviside $\Theta$ factors in $\eta_{ij}(E_n, E')$.

\Cref{eq:lindblad-general,eq:lindop} capture the physics of the formalism described in \cref{sec:existing-results} (though in much more easily generalizable way), under the approximation that the final state neutrino energy is discretized, i.e.\ fixed at the corresponding bin centre. The oscillation phase and decay exponential, $\upphi_{i,j}$, are accounted for naturally by the evolution of the ODE.

We solve the Lindblad equation in \texttt{Python} using the \texttt{odeintw} \cite{odeintw} package, based on the LSODA algorithm from \texttt{odepack} \cite{odepack} as implemented in \texttt{SciPy} \cite{scipy}. Any equivalent matrix ODE solver in any programming language could also be employed.

\begin{figure}
    \centering
    \includegraphics[width=\linewidth]{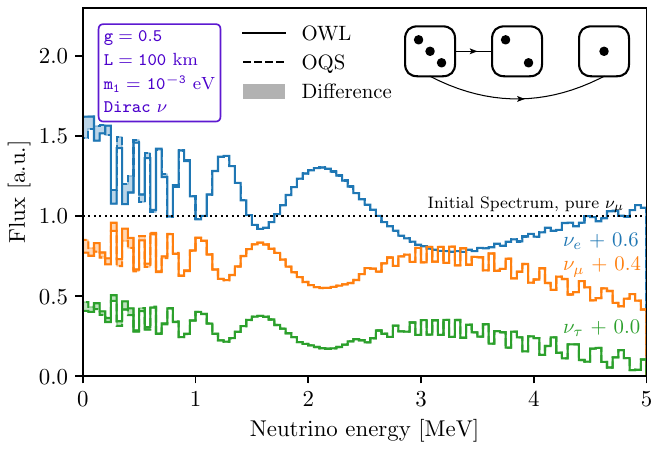}
    \caption{Neutrino decay in a simple toy scenario with three neutrino flavours and only two decay modes, $\nu_3 \to \nu_1$ and $\nu_3 \to \nu_2$. We compare the phenomenological approach due to Ohlsson, Winter, and Lindner (OWL) \cite{Lindner:2001fx} to the open quantum system (OQS) approach. In the latter case, solving the Lindblad equation or applying Kraus operators gives identical results. The minor differences between the OWL and OQS methods are due to energy discretization effects in the OQS approach. A pure $\nu_\mu$ initial flux with a flat spectrum from \SIrange{0}{5}{MeV} is assumed (dotted black line), and we plot the final flux at a distance of \SI{100}{km}. We have used $m_1 = \SI{e-3}{eV}$.}
    \label{fig:check}
\end{figure}

In \cref{fig:check} we compare the flavour transition probabilities calculated in the OWL and Lindblad approaches for a simple toy scenario with three neutrino flavours and the decay modes $\nu_3 \to \nu_1$ and $\nu_3 \to \nu_2$. This scenario has been considered previously for instance in ref.~\cite{Coloma:2017zpg}. For more complex scenarios, such as models in which also $\nu_2 \to \nu_1$ occurs, or in which there are more than three neutrino flavours, the OWL approach quickly becomes fairly complex \cite{Lindner:2001fx, Abdullahi:2020rge}, while the Lindblad formalism remains compact.  We assume a pure $\nu_\mu$ initial state with a flat initial spectrum. For illustrative purposes, we choose unrealistically large couplings $g_{ij} = 0.5$ and a baseline of \SI{100}{km} -- long enough for decays to occur, but short enough for oscillations not to average out.

In terms of observable predictions, the two formalisms agree, of course. The small differences visible in \cref{fig:check} are due to the necessary discretization of energy in the Lindblad approach: when a neutrino has decayed, the Lindblad equation places the daughter particle at the centre of its energy bin (and subsequently oscillates it at that energy). In contrast, the unbinned OWL approach retains the exact daughter energy.

\begin{figure}
    \centering
    \includegraphics[width=0.46\textwidth]{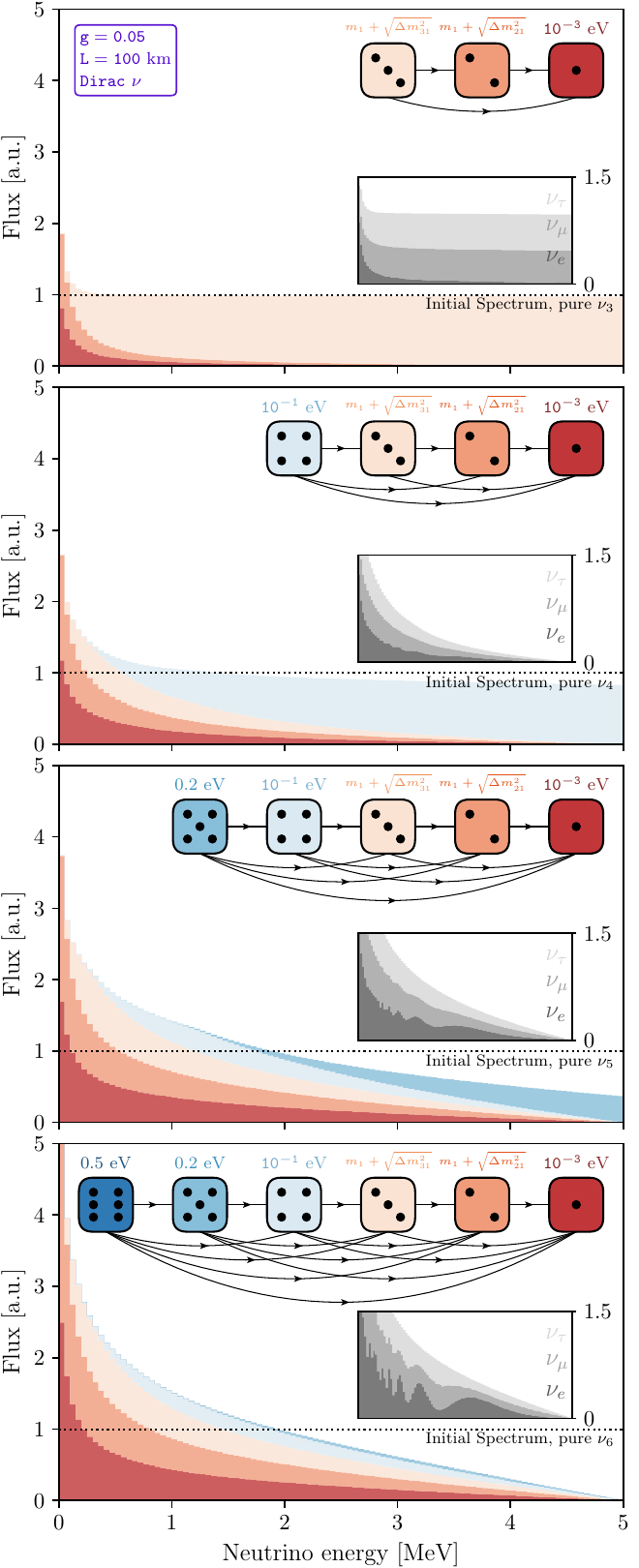}
    \caption{Neutrino decay in complex scenarios with $\geq 3$ neutrino flavours and with multiple decay modes as indicated in each panel. In all cases, we assume an initial flux consisting purely of the heaviest mass eigenstate, and with a flat spectrum from \SIrange{0}{5}{MeV} (dotted horizontal lines). The composition of the final state flux in the mass basis is shown as a stacked histogram.  We choose a baseline $L=\SI{50}{km}$, and neutrino masses as indicated in the plots. The greyscale inlay in each panels presents the results for the three SM neutrinos in the flavour basis. (We assume no active--sterile neutrino mixing for simplicity.)}
    \label{fig:lindblad-show-off}
\end{figure}

As mentioned above, the most important advantage of the Lindblad approach over the OWL method is that it can be straightforwardly applied to arbitrarily complex systems with multiple decay modes and possibly more than three neutrino flavours, without adding significant complexity. To illustrate this, we show in \cref{fig:lindblad-show-off} results for toy systems with up to six neutrino species and up to 15 decay modes, including cascades. Results are shown both in the mass basis (main plots, stacked coloured histograms) and in the flavour basis (grey insets).

\begin{figure}
    \vspace{-0.2cm}
    \centering
    \includegraphics[width=\linewidth]{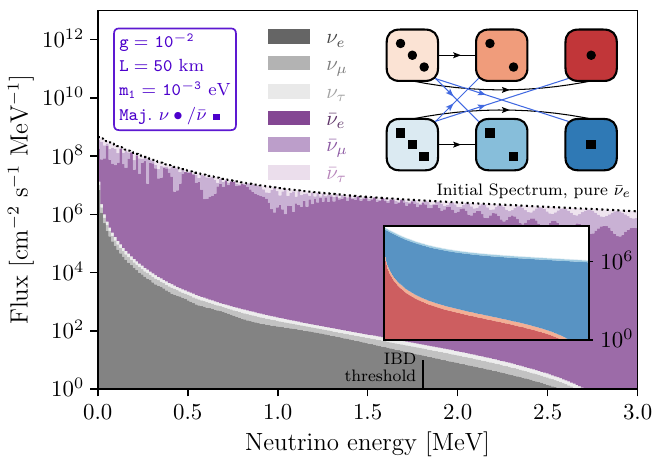}
    \caption{Neutrino decay in a more realistic picture, with a pure $\bar{\nu}_e$ flux from a nuclear reactor at a \SI{50}{km} distance as in JUNO \cite{JUNO:2021vlw} or KamLAND \cite{KamLAND:2008dgz}. We have one unstable neutrino, and we allow for the possibility of helicity-violating decays (see relevant formulae in \cref{sec:decay}.}
    \label{fig:majorana}
\end{figure}

Finally, in \cref{fig:majorana}, we illustrate a more realistic scenario inspired by ref.~\cite{JUNO:2021ydg}: we consider the $\bar\nu_e$ flux from a nuclear reactor, observed at a distance of \SI{50}{km} in a JUNO-like experiment \cite{JUNO:2021vlw}. Unlike in the previous plots, we assume here that neutrinos are Majorana particles, so that both lepton number conserving and lepton number violating decays are allowed. This requires us to track six neutrino species (three neutrino species, three anti-neutrino species) with four possible decay modes. The density matrix in flavour space therefore takes the form
\begin{align}
    \rho = \begin{pmatrix}
               \rho_\nu               & \rho_{\nu\bar\nu} \\
               \rho_{\nu\bar\nu}^\dag & \rho_{\bar{\nu}}
            \end{pmatrix},
\end{align}
where the first three rows and columns correspond to neutrinos, while rows and columns 4--6 represent anti-neutrinos. The off-diagonal blocks encode correlations between neutrino and anti-neutrino modes. Note that we do not include $\parenbar\nu_2 \to \parenbar\nu_1$ decays here because of stringent existing constraints on the corresponding couplings, which render the $\parenbar\nu_2$ lifetime unobservably long. The $\parenbar\nu_3$ decays are of course also constrained experimentally. The scenarios studied in ref.~\cite{JUNO:2021ydg} and \cref{fig:majorana} are consistent with terrestrial constraints, but disfavoured by constraints from IceCube and from the CMB \cite{Abdullahi:2020rge, Escudero:2019gfk} unless there is additional new physics to circumvent those constraints.

We see in \cref{fig:majorana} how our methods have no difficulty tracking this complicated system, with the final spectra exhibiting the expected oscillatory pattern as well as the build-up of daughter neutrinos and anti-neutrinos, especially at low energies.

\subsection{Dynamical Map and Kraus Operators}
\label{sec:kraus}

While the Lindblad formalism provides a fully general, flexible, and efficient description of neutrino decay and oscillations, it is often not optimal from a computational point of view, given that it still requires solving a differential equation. An example is the evolution over very long time intervals. This can be avoided by utilising the dynamical map or Kraus operators (cf.\ \cref{sec:general-formalism}) which directly encode the time evolution over arbitrarily long intervals. As we have seen in \cref{sec:general-formalism}, we thereby replace integrating a differential equation with exponentiating a matrix and solving an eigenvector problem.

More precisely, consider again an $N_\nu$-flavour ensemble of neutrinos distributed across $N_E$ energy bins. As in \cref{sec:lindblad-approach}, the system can be described by an $N_\nu N_E \times N_\nu N_E$ density matrix, though only the $N_E$ diagonal $N_\nu \times N_\nu$ blocks are relevant in practice, given that correlations between neutrinos with macroscopically different energies decohere instantaneously.

From the $N_\nu N_E \times N_\nu N_E$ density matrix, we can construct the $(N_\nu N_E)^2 \times (N_\nu N_E)^2$ Liouvillian superoperator, $\hat{\mathcal{L}}$, using \cref{eq:ume}. The resulting dynamical map $\mathcal{E}(L) = e^{\hat{\mathcal{L}}L}$ has the same dimension.  $\mathcal{E}(L)$ formally solves the system (cf.\ \cref{eq:t-evolution-dynamical-map}):
\begin{align}
    \operatorname{vec}(\rho(L)) = \mathcal{E}(L) \, \operatorname{vec}(\rho(0)).
    \label{eq:unrestricted-rho-solution}
\end{align}
However, as we are only interested in the diagonal $N_\nu \times N_\nu$ blocks of the density matrix, we can simplify the calculation by unravelling only the diagonal blocks of $\rho$ and ignoring the off-diagonal ones. We define the \emph{stacked vectorization} of $\rho$ as
\begin{align}
    \operatorname{vec}(\tilde{\rho}) =
    \begin{pmatrix}
        \operatorname{vec}(\rho^{(1)}) \\
        \operatorname{vec}(\rho^{(2)}) \\
        \vdots \\
        \operatorname{vec}(\rho^{(N_E)})
    \end{pmatrix} ,
\end{align}
where $\rho^{(1)}, \ldots, \rho^{(N)}$ denote the diagonal $N_\nu \times N_\nu$ blocks of $\rho$. We can also restrict the Liouvillian superoperator $\hat{\mathcal{L}}$ to the subspace of $\operatorname{vec}(\tilde{\rho})$. Denoting the thus restricted Liouvillian $\tilde{\hat{\mathcal{L}}}$ and using as before indices $n$, $m$ to enumerate energy bins, the $N_\nu^2 \times N_\nu^2$ blocks of $\tilde{\hat{\mathcal{L}}}$ are
\begin{align}
    \tilde{\hat{\mathcal{L}}}^{(nn)}
        &= -i \left( \mathbb{I} \otimes H^{(n)} - (H^{(n)})^T \otimes \mathbb{I} \right) \notag\\ 
        &- \frac{1}{2} \sum_m \Big(
               \mathbb{I} \otimes L^{(nm)\dag} L^{(nm)}
             + [L^{(nm)\dag} L^{(nm)}]^T \otimes \mathbb{I}
           \Big) \\
    \tilde{\hat{\mathcal{L}}}^{(nm)}
        &= L^{*(mn)} \otimes L^{(mn)}, \qquad\qquad n \neq m.
\end{align}
Exponentiating yields the dynamical map:
\begin{align}
    \tilde{\mathcal{E}}(L) = e^{\tilde{\hat{\mathcal{L}}} L}
        = \begin{pmatrix}
              \mathcal{E}^{(00)} & \mathcal{E}^{(01)} & \cdots \\
              \mathcal{E}^{(10)} & \mathcal{E}^{(11)} & \cdots \\
        \vdots & & \ddots
    \end{pmatrix}.
    \label{eq:dynamical-map-simplified}
\end{align}
This operator -- an $N_\nu^2 N_E \times N_\nu^2 N_E$ matrix -- explicitly solves the system by allowing us to compute directly
\begin{align}
    \operatorname{vec}(\tilde{\rho}(L))
        = \tilde{\mathcal{E}}(L) \, \operatorname{vec}(\tilde{\rho}(0)).
    \label{eq:restricted-rho-solution}
\end{align}
Undoing the unravelling operation in $\operatorname{vec}(\tilde{\rho}(L))$, we recover the diagonal blocks of $\rho(L)$. Therefore, \cref{eq:restricted-rho-solution} offers a computationally more efficient way of computing these diagonal blocks than working with full $(N_\nu N_E)^2 \times (N_\nu N_E)^2$ matrices as in \cref{eq:unrestricted-rho-solution}.

Since each block $\mathcal{E}^{(mn)}(L)$ of $\tilde{\mathcal{E}}$ is an $N_\nu^2 \times N_\nu^2$ superoperator, we can construct a set of Kraus operators $\{M_k^{(mn)}\}$ of dimension $N_\nu \times N_\nu$ from each of them. In other words, we carry out the steps outlined in \cref{eq:shuffle,eq:Choi-decomposition,eq:Choi-decomposition-components,eq:t-evolution-kraus1,eq:kraus-op-definition} for each $\mathcal{E}^{(mn)}(L)$ block separately. Unpacking $\operatorname{vec}(\tilde{\rho}(L))$ bin-by-bin yields the evolution equation for the neutrinos in the $m$-th energy bin,
\begin{align}
    \rho^{(m)}(L) = \sum_{n=1}^{N_E} \sum_k M_k^{(mn)}\,\rho^{(n)}(0) \, M_k^{(mn)\dagger},
\end{align}
where the sum over $k$ runs over the set of Kraus operators corresponding to $\mathcal{E}^{(mn)}(L)$, and the sum over energy bins $n$ accounts for the off-diagonal blocks of $\tilde{\mathcal{E}}(L)$---the population in $\rho^{(m)}(L)$ receives contributions from all initial bins $m$ that decay to $n$, via the Lindblad operators $L_{nm}$.

We have implemented the optimized derivation of the Kraus operators outlined here in the public code package accompanying this paper \cite{github}, and we have verified that its results agree with those obtained by solving the Lindblad equation.

\vspace{1.5ex} \noindent
For many simple systems, the Kraus operators can be inferred more directly from physical arguments. In the following, we give some examples, starting from the simplest toy model (a single decay branch without oscillation) and moving towards more general and realistic systems. We assume here only three neutrino flavours.

\subsubsection{Single Decay}
\label{sec:single-decay}

\begin{figure}[H]
    \centering
    \includegraphics[]{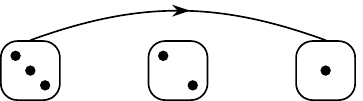}
\end{figure}

For a single decay, $\nu_3 \to \nu_1$, there are $N_E (N_E+1) / 2$ Lindblad operators, each corresponding to one block of \cref{eq:dynamical-map-simplified}, that is, to decays from a particular initial energy bin, $n$, to a particular final energy bin, $m$, with $n \geq m$. The $(m,n)$ block of each Lindblad operator will have the structure
\begin{align}
    \begin{pmatrix}
        0 & 0 & \bsq \\
        0 & 0 & 0    \\
        0 & 0 & 0
    \end{pmatrix}.
\end{align}
(with all other blocks being zero). In the absence of oscillations, this system is almost identical to the toy example of a radiatively decaying atomic excitation introduced in \cref{sec:toy-example}. Consequently, the form of the Kraus operators can be read off directly from \cref{eq:kraus-operators-toy}:
\begin{align}
    \begin{split}
        M_0^{(mn)\cancel{\text{osc}}}(L) &= \delta_{nm} \begin{pmatrix}
                      1 & 0 & 0 \\
                      0 & 1 & 0 \\
                      0 & 0 & \sqrt{e^{-\Gamma_{31} L}}
                  \end{pmatrix}, \\
        M_1^{(mn)\cancel{\text{osc}}}(L) &= \gamma_{31}^{(nm)} \begin{pmatrix}
                      0 & 0 & \sqrt{\Gamma_{31}^{-1} (1 - e^{-\Gamma_{31}})} \\
                      0 & 0 & 0 \\
                      0 & 0 & 0\end{pmatrix} \!.
    \end{split}
    \label{eq:G31_noH}
\end{align}
where the $\upgamma_{31}^{(nm)}$ have been defined in \cref{eq:gamma-ijnm} as the square root of the differential decay rate, $\eta_{ij}$, over the $m$-th energy bin. For later use, it will be useful to define the decay integral
\begin{align}
    \Iint{A}{B} 
      = \int_0^L e^{-{\color{ingoing}A}\,l}\, e^{-{\color{outgoing}B}(L-l)}\, dl
      = \frac{e^{-{\color{outgoing}B} L} - e^{-{\color{ingoing}A} L}}
             {{\color{ingoing}A} - {\color{outgoing}B}} \,,
\end{align}
where $\color{ingoing}A$ and $\color{outgoing}B$ are the accumulated phases of the ingoing and outgoing neutrinos, respectively. This allows us to write $M_1^{(mn)\cancel{\text{osc}}}(L)$ in a more compact way
\begin{align}
    M_1^{(mn)\cancel{\text{osc}}}(L) &= \gamma_{31}^{(nm)} \begin{pmatrix}
                  0 & 0 & \sqrt{\Iint{\Gamma_{31}}{0}} \\
                  0 & 0 & 0 \\
                  0 & 0 & 0\end{pmatrix} \!.
\end{align}
Reintroducing oscillations modifies only the diagonal Kraus operators $M_0^{(mn)\cancel{\text{osc}}}(L)$ to
\begin{multline}
    M_0^{(mn)}(L) = \delta_{nm} \operatorname{diag}\bigg(1,\,
    \,\exp\bigg[-i\frac{\Delta m_{21}^2}{2E_n} L \bigg], \, \\
    \,\exp\bigg[ \bigg( -i \frac{\Delta m_{31}^2}{2E_n}
        - \frac{\alpha_{31}}{2E_n} \bigg) L \bigg] \bigg) \,.
    \label{eq:G31}
\end{multline}

\subsubsection{Dual Decay}

\begin{figure}[H]
    \centering
    \includegraphics[]{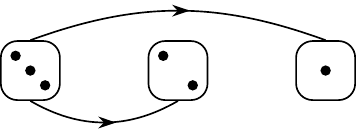}
\end{figure}

Consider now multiple decays from a single unstable neutrino $\nu_3 \to \nu_2, \,\,\, \nu_3 \to \nu_1$. Now, each of the $N_E (N_E + 1) / 2$ Lindblad operators will have the form
\begin{align}
    \begin{pmatrix}
        0 & 0 & \bsq \\
        0 & 0 & \bsq \\
        0 & 0 & 0
    \end{pmatrix}.
\end{align}
By similar arguments as in \cref{sec:single-decay}, the form of the Kraus operators in the absence of oscillations can again be inferred:
\begin{align}
    \begin{split}
        M_0^{(mn)\cancel{\text{osc}}}(L) &= \delta_{nm} \begin{pmatrix}
                      1 & 0 & 0 \\
                      0 & 1 & 0 \\
                      0 & 0 & \sqrt{e^{-\Gamma_3 L}}
                  \end{pmatrix}, \\[0.2cm]
        M_1^{(mn)\cancel{\text{osc}}}(L) &= \begin{pmatrix}
                      0 & 0 & \gamma_{31}^{(nm)}\sqrt{\Gamma_{31}\,\Iint{\Gamma_{3}}{0}} \\
                      0 & 0 & \gamma_{32}^{(nm)}\sqrt{\Gamma_{32}\,\Iint{\Gamma_{3}}{0}} \\
                      0 & 0 & 0\end{pmatrix} \!.
    \end{split}
    \label{eq:double-decay-kraus-noosc}
\end{align}
where $\Gamma_3 \equiv \Gamma_{31} + \Gamma_{32}$ is the total $\nu_3$ decay rate. It will be useful to look at the structure of the dynamical map from which these Kraus operators are derived. Putting back oscillations, each of the blocks in \cref{eq:dynamical-map-simplified} -- which for $N_\nu = 3$ are $9 \times 9$ matrices -- has the structure
\begin{align}
    \begin{pmatrix}
        \gsq & 0    & 0    & 0    & 0    & 0    & 0    & 0    & \bsq \\
        0    & \gsq & 0    & 0    & 0    & 0    & 0    & 0    & \bsq \\
        0    & 0    & \gsq & 0    & 0    & 0    & 0    & 0    & 0    \\
        0    & 0    & 0    & \gsq & 0    & 0    & 0    & 0    & \bsq \\
        0    & 0    & 0    & 0    & \gsq & 0    & 0    & 0    & \bsq \\
        0    & 0    & 0    & 0    & 0    & \gsq & 0    & 0    & 0    \\
        0    & 0    & 0    & 0    & 0    & 0    & \gsq & 0    & 0    \\
        0    & 0    & 0    & 0    & 0    & 0    & 0    & \gsq & 0    \\
        0    & 0    & 0    & 0    & 0    & 0    & 0    & 0    & \gsq
    \end{pmatrix}.
\end{align}
Elements shown in grey here are the ones describing oscillations and disappearance of parent neutrinos. These elements are non-zero only in the diagonal blocks of \cref{eq:dynamical-map-simplified}. Elements shown in black are the ones describing decay. Rearranging the dynamical map into the Choi matrix according to \cref{eq:Choi-decomposition-components}, we infer that the Choi matrix has the structure
\begin{align}
    \begin{pmatrix}
        \gsq & 0    & 0    & 0    & \gsq & 0    & 0    & 0    & \gsq \\
        0    & 0    & 0    & 0    & 0    & 0    & 0    & 0    & 0    \\
        0    & 0    & 0    & 0    & 0    & 0    & 0    & 0    & 0    \\
        0    & 0    & 0    & 0    & 0    & 0    & 0    & 0    & 0    \\
        \gsq & 0    & 0    & 0    & \gsq & 0    & 0    & 0    & \gsq \\
        0    & 0    & 0    & 0    & 0    & 0    & 0    & 0    & 0    \\
        0    & 0    & 0    & 0    & 0    & 0    & \bsq & \bsq & 0    \\
        0    & 0    & 0    & 0    & 0    & 0    & \bsq & \bsq & 0    \\
        \gsq & 0    & 0    & 0    & \gsq & 0    & 0    & 0    & \gsq
    \end{pmatrix}.
\end{align}
The decay terms now form a $2 \times 2$ block which reads
\begin{align}
    \Iint{\Gamma_{3}}{0}
    \begin{pmatrix}
        \gamma^{(nm)}_{31} & \big[\gamma^{(nm)}_{31} \gamma^{(nm)}_{32}\big]^\frac12 \\ 
        \big[\gamma^{(nm)}_{31} \gamma^{(nm)}_{32}\big]^\frac12 & \gamma^{(nm)}_{32}
    \end{pmatrix}.
    \label{eq:double-decay-subblock-noosc}
\end{align}
without oscillations, and
\begin{align}
    \begin{pmatrix}
        \gamma^{(nm)}_{31} \, \Iint{\Gamma_{3}}{0}
            & \!\!\!\!\!\big[\gamma^{\!(nm)}_{31} \! \gamma^{\!(nm)}_{32}\big]^\frac12 \,
              \!\Iint{\!\Gamma_{3}}{\! -\frac{i \Delta m_{21}^2}{2E_m}\!}\!\! \\[3pt]
        \!\big[\gamma^{\!(nm)}_{31} \! \gamma^{\!(nm)}_{32}\big]^\frac12 \,
            \!\Iint{\!\Gamma_{3}}{\!\frac{i \Delta m_{21}^2}{2E_m}\!}
            & \gamma^{(nm)}_{32} \, \Iint{\Gamma_{3}}{0}
    \end{pmatrix}
    \label{eq:double-decay-subblock-osc}
\end{align}
with oscillations included. The eigenvalues and eigenvectors of this $2 \times 2$ subblock determine the non-diagonal Kraus operators. Without oscillations, \cref{eq:double-decay-subblock-noosc} has rank~1, so there is only one non-diagonal Kraus operators, given by \cref{eq:double-decay-kraus-noosc}. With oscillations, there are two non-diagonal Kraus operators. (As the corresponding expressions are more lengthy, we do not give them explicitly here.)

\subsubsection{Cascade Decay}

\begin{figure}[H]
    \centering
    \includegraphics[]{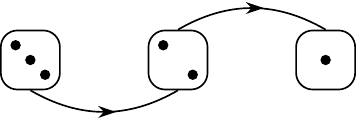}
\end{figure}

Consider now a decay chain $\nu_3 \to \nu_2 \to \nu_1$, for which the $N_E (N_E+1) / 2$ pairs of Lindblad operators take the form
\begin{align}
    \begin{pmatrix}
        0 & 0 & 0    \\
        0 & 0 & \bsq \\
        0 & 0 & 0
    \end{pmatrix}.
    \qquad
    \begin{pmatrix}
        0 & \bsq & 0    \\
        0 & 0    & 0 \\
        0 & 0    & 0
    \end{pmatrix}.
\end{align}
The first matrix drives  $\nu_3 \to \nu_2$ transitions and the second one drives $\nu_2 \to \nu_1$ decays. The blocks of the dynamical map $\mathcal{E}(L)$ now have the structure
\begin{align}
    \begin{pmatrix}
        \gsq & 0    & 0    & 0    & \bsq & 0    & 0    & 0    & \bsq \\
        0    & \gsq & 0    & 0    & 0    & 0    & 0    & 0    & 0    \\
        0    & 0    & \gsq & 0    & 0    & 0    & 0    & 0    & 0    \\
        0    & 0    & 0    & \gsq & 0    & 0    & 0    & 0    & 0    \\
        0    & 0    & 0    & 0    & \gsq & 0    & 0    & 0    & \bsq \\
        0    & 0    & 0    & 0    & 0    & \gsq & 0    & 0    & 0    \\
        0    & 0    & 0    & 0    & 0    & 0    & \gsq & 0    & 0    \\
        0    & 0    & 0    & 0    & 0    & 0    & 0    & \gsq & 0    \\
        0    & 0    & 0    & 0    & 0    & 0    & 0    & 0    & \gsq
    \end{pmatrix},
\end{align}
where, as in the previous section, the grey elements are related to the diagonal Kraus operator, while the black ones describe how the $(2,2)$ and $(1,1)$ elements of the density matrix are populated at the expense of the $(3,3)$ and $(2,2)$ elements, respectively, and how the $(1,2)$ / $(2,3)$ and $(2,1)$ / $(3,2)$ elements get correlated. Translated to a Choi matrix, we obtain the structure:
\begin{align}
    \begin{pmatrix}
        \gsq & 0    & 0    & 0    & \gsq & 0    & 0    & 0    & \gsq \\
        0    & 0    & 0    & 0    & 0    & 0    & 0    & 0    & 0    \\
        0    & 0    & 0    & 0    & 0    & 0    & 0    & 0    & 0    \\
        0    & 0    & 0    & \bsq & 0    & 0    & 0    & 0    & 0    \\
        \gsq & 0    & 0    & 0    & \gsq & 0    & 0    & 0    & \gsq \\
        0    & 0    & 0    & 0    & 0    & 0    & 0    & 0    & 0    \\
        0    & 0    & 0    & 0    & 0    & 0    & \bsq & 0    & 0    \\
        0    & 0    & 0    & 0    & 0    & 0    & 0    & \bsq & 0    \\
        \gsq & 0    & 0    & 0    & \gsq & 0    & 0    & 0    & \gsq  
    \end{pmatrix},
\end{align}
The $2 \times 2$ block in black (4th and 8th row and column) has the structure
\begin{align}
    \begin{pmatrix}
        \gamma_{21}^{(nm)} \Iint{\Gamma_{21}^{(n)}}{0} & 0 \\[6pt]
        0 & \gamma_{32}^{(nm)} \Iint{\Gamma_{32}}{\Gamma_{21}^{(m)}}
    \end{pmatrix} .
    \label{eq:cascade-22subblock}
\end{align}
The element in the $(7,7)$ position encoding the $\nu_3 \to \nu_2 \to \nu_1$ cascade is given by
\begin{align}
    \mathcal{C}_{321}
        &\equiv \sum_p \gamma_{32}^{(np)} \, \gamma_{21}^{(pm')} \!
            \int_0^L\! dl_2\! \int_0^{l_2}\! dl_1\, e^{-\Gamma_{32} l_1}
                                                    e^{-\Gamma_{21}^{(p)}(l_2 - l_1)}
                                                                \notag\\
        &= \sum_p \frac{\gamma_{32}^{(np)} \, \gamma_{21}^{(pm')}}
                       {\Gamma_{32} - \Gamma_{21}^{(p)}} \bigg(
               \frac{1 - e^{-\Gamma_{21}^{(p)} L}}{\Gamma_{21}^{(p)}}
             - \frac{1-e^{-\Gamma_{32} L}}{\Gamma_{32}} \bigg) .
    \label{eq:cascade_int}
\end{align}
From these considerations, one can derive the Kraus operators, which we give here only for the case without oscillations for brevity. The diagonal Kraus operator is
\begin{align}
    M_0^{(mn)\cancel{\text{osc}}}(L) = \delta_{nm}
        \begin{pmatrix}
            1 & 0 & 0 \\
            0 & \sqrt{e^{-\Gamma_{21} L}} & 0 \\
            0 & 0 & \sqrt{e^{-\Gamma_{32} L}}
        \end{pmatrix},
\end{align}
and the Kraus operator describing the cascade, derived from \cref{eq:cascade_int}, is
\begin{align}
    M_1^{(mn)\cancel{\text{osc}}}(L) =
        \begin{pmatrix}
            0 & 0 & \sqrt{\mathcal{C}_{321}} \\
            0 & 0 & 0 \\
            0 & 0 & 0
        \end{pmatrix}.
\end{align}
There are two additional Kraus operators, corresponding to the eigenvalues $\lambda_k$ and the eigenvectors $\vec{v}_k = (v_{k1}, v_{k2})$ ($k=1,2$) of \cref{eq:cascade-22subblock}. These operators have the form
\begin{align}
    M_{k+1}^{(nm)\cancel{\text{osc}}}(L) = \sqrt{\lambda_k}
        \begin{pmatrix}
            0 & v_{k1} & 0 \\
            0 & 0 & v_{k2} \\
            0 & 0 & 0
        \end{pmatrix}, \quad k=1,2.
\end{align}

\section{Discussion}
\label{sec:discussion}

\begin{figure}[H]
    \centering
    \includegraphics[width=\linewidth]{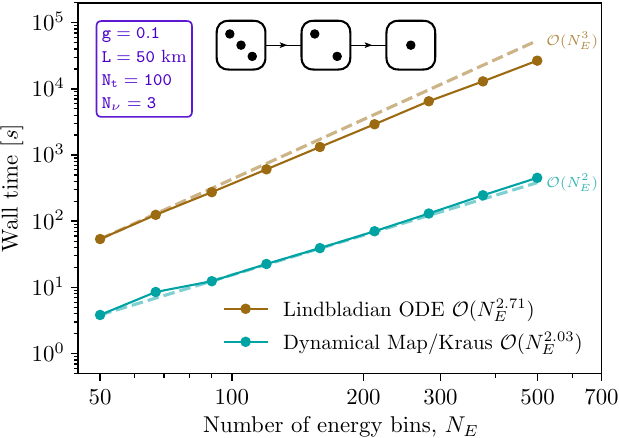}
    \caption{Numerical complexity of different algorithms for computing combined neutrino oscillation+decay probabilities as a function of the number of energy bins, $N_E$. Solid lines indicate the empirical wall clock time, while dashed lines show the theoretical scaling.}
    \label{fig:complexity}
\end{figure}

\begin{table}
    \centering
    \begin{tabularx}{\linewidth}{lll}
        \toprule
                     & Complexity & Remarks \\
        \midrule
        \bf OWL      & $\mathcal{O}\big[N_\nu N_E N_t \big]$      &
                                for simple systems \\[0.1cm]
        \bf Lindblad ODE & $\mathcal{O}\big[(N_\nu N_E)^3 N_t \big]$  &
                                general-purpose \\[0.1cm]
        \bf Dyn.\,Map/Kraus & $\mathcal{O}\big[N_\nu^6 N_E^{2\text{--}3} \big]$ &
                                most efficient, no ODE \\
        \bottomrule
    \end{tabularx}
    \caption{Comparison between the different approaches to the neutrino oscillation+decay problem discussed in this paper. For each formalism, we indicate the computational complexity as a function of the number of neutrino species, $N_\nu$, the number of energy bin, $N_E$, and the number of time/distance steps, $N_t$.}
    \label{tab:complexity}
\end{table}

We now compare the different approaches to the neutrino oscillation+decay problem discussed in this paper.

{\bf Direct analytic calculations} (which we dubbed OWL for Ohlsson, Winter, Lindner) are an excellent choice for simple systems ($\leq 3$ neutrino flavours, no multi-step decays). Their computational complexity scales proportional to $N_\nu$ (number of neutrino flavours), $N_E$ (number of energy bins), and $N_t$ (number of time steps). For simple systems (few relevant neutrino flavours, only one or two decay modes, no multi-step decays), the time integral can be easily carried out analytically, allowing for a fully analytic solution. For more complex systems, the analytic expressions become extremely lengthy and therefore costly to evaluate. Moreover, it is very difficult to prevent numerical errors from spinning out of control when evaluating such lengthy expressions.

The {\bf Lindblad approach} remains conceptually simple even for very complex systems, but it requires solving an ordinary differential equation, which usually needs to be done numerically. In terms of computational complexity, the evaluation of the right-hand side of the Lindblad equation requires multiplication of $(N_\nu N_E) \times (N_\nu N_E)$ matrices, with a complexity of $\mathcal{O}\big( (N_\nu N_E)^3 \big)$. The effort for evolving the system in time is at least proportional to $N_t$, entailing a total complexity of $\mathcal{O}\big( (N_\nu N_E)^3 N_t \big)$%
\footnote{This is obviously true when using an explicit ODE solver, which requires one evaluation of the right-hand side in each time step. It is, however, true even for implicit solvers. Such solvers typically require solving a system of linear equations of the form $\rho_{n+1} + dt \, (L^\dag \rho L - \frac{1}{2} \{ L^\dag L, \rho_{n+1} \} = \rho_n$, which has complexity $\mathcal{O}\big( (N_\nu N_E)^3 N_t \big)$. Note also that $N_t$ does not scale linearly with the baseline, $L$. At large $L$, numerical errors begin to accumulate, and to counteract them a typical ODE solver will choose smaller step sizes. Note also that at large $L$, the number of energy bins, $N_E$, needs to be increased to avoid aliasing artifacts, unless some energy smearing is introduced as a regulator, for instance through additional decoherence-like Lindblad operators.}
We consider the Lindblad approach a useful general purpose method as it can straightforwardly handle arbitrarily complex neutrino decay problems. Moreover, it is easily expandable if additional effects are to be taken into account, for instance neutrino absorption or decoherence.

{\bf The dynamical map and Kraus operators} naively seem to come with increased complexity as their derivation requires exponentiation of an $(N_\nu N_E)^2 \times (N_\nu N_E)^2$ matrix. Whether the latter operation is carried out via diagonalization of the matrix or via evaluating a truncated exponential series, the complexity is $\mathcal{O}\big[(N_\nu N_E)^6 \big]$ in either case. As we have seen in \cref{sec:kraus}, however, the complexity can be significantly reduced by exploiting the fact that the off-diagonal $N_\nu \times N_\nu$ blocks of the $N_\nu N_E \times N_\nu N_E$ density matrix are not needed. This reduces the computational complexity to $\mathcal{O}\big(N_\nu^6 N_E^3 \big)$, where the scaling with $N_E$ stems from the complexity of the matrix exponentiation needed to compute the dynamical map from the Liouvillian superoperator. In practice, it turns out that the off-diagonal elements of the Liouvillian superoperator are often significantly smaller than the diagonal ones, and moreover the off-diagonal blocks are sparsely populated if only certain decay modes are relevant. We find that this improves the scaling to $\sim \mathcal{O}\big(N_\nu^6 N_E^2 \big)$. The Kraus operator approach is therefore the fastest general-purpose method (unless $N_\nu$ is \emph{very} large, $\gg 3$). In addition, Kraus operators directly \emph{solve} the system for arbitrary $L$, making it even more advantageous when the evolution needs to be tracked over very long distances, or with very small time steps.\footnote{Note, though, that because the neutrino oscillation phase is proportional to $L/E$, evolution over long distances typically also requires many energy bins in order not to lose resolution. To avoid this, extra decoherence terms can be added to explicitly smear out fast oscillations at low energy/long distance.}

The computational complexity of the different methods discussed in this paper is summarized in \cref{tab:complexity,fig:complexity}

\section{Conclusions}
\label{sec:conclusions}

In summary, we have investigated the theory of neutrino oscillation and decay. By applying methods from the theory of open quantum systems, we have found that arbitrarily complicated oscillation+decay problems can be tackled with high numerical efficiency, either by solving the Lindblad master equation or by computing a set of Kraus operators which fully solve the system without the need for integrating a differential equation. Finally, we have shown that in some cases, these methods admit closed-form solutions, which capture the intricate physics of these systems in a form inspired by quantum information theory.

We hope that these results, and the Python package accompanying this paper \cite{github}, will be useful to both the theoretical and experimental neutrino communities for future studies of extensions of the Standard Model involving decaying active or sterile neutrinos. Excited by the gamut of precision neutrino oscillation experiments, operating at the present and near-future, we look forward to the possibility of \emph{unravelling} the quantum secrets of neutrinos.

\begin{acknowledgments}
\noindent
The authors gratefully acknowledge many helpful discussions, across time and space, including with Leonardo Ferreira Leite, Gustavo Alves, and Carlos Arg\"uelles. We also thank Michael Wurm for continued discussions and support, and the wider JUNO Collaboration for their interest in this work. GAP acknowledges support from the DFG Research Unit FOR~5519, \emph{Precision Neutrino Physics in JUNO}. JK and GAP are supported by the Cluster of Excellence \emph{Precision Physics, Fundamental Interactions and Structure of Matter} (PRISMA++, EXC 2118/2, Project ID 390831469), funded by the German Research Foundation (DFG).
\end{acknowledgments}

\bibliography{refs}

\end{document}